\documentclass{article}

\PassOptionsToPackage{numbers}{natbib}

    \usepackage[preprint]{neurips_2025}

\usepackage[utf8]{inputenc} 
\usepackage[T1]{fontenc}    
\usepackage{hyperref}       
\usepackage{url}            
\usepackage{booktabs}       
\usepackage{amsfonts}       
\usepackage{nicefrac}       
\usepackage{microtype}      
\usepackage{xcolor}         
\usepackage{graphicx}
\usepackage{subcaption} 
\usepackage{multirow}

\title{HPP-Voice: A Large-Scale Evaluation of Speech Embeddings for Multi-Phenotypic Classification}

\author{%
  David Krongauz\thanks{Corresponding author.} \\
  \texttt{david.krongauz@weizmann.ac.il} \\
  \And
  Hido Pinto \\
  \texttt{yehudahido.pinto@weizmann.ac.il} \\
  \And
  Sarah Kohn \\
  \texttt{sarah.kohn@weizmann.ac.il} \\
  \And
  Yanir Marmor \\
  \texttt{yanir.marmor@weizmann.ac.il} \\
  \And
  Eran Segal \\
  \texttt{eran.segal@weizmann.ac.il} \\
  \\
  Department of Computer Science and Applied Mathematics,\\
  Department of Molecular Cell Biology,\\
  Weizmann Institute of Science, Rehovot, Israel \\
}

\begin{document}

\maketitle

\begin{abstract}
Human speech contains paralinguistic cues that reflect a speaker's physiological and neurological state, potentially enabling non-invasive detection of various medical phenotypes. We introduce the Human Phenotype Project Voice corpus (HPP-Voice): a dataset of 7,188 recordings in which Hebrew-speaking adults count for 30 seconds, with each speaker linked to up to 15 potentially voice-related phenotypes spanning respiratory, sleep, mental health, metabolic, immune, and neurological conditions. We present a systematic comparison of 14 modern speech embedding models, where modern speech embeddings from these 30-second counting tasks outperform MFCCs and demographics for downstream health condition classifications. We found that embedding learned from a speaker identification model can predict objectively measured moderate to severe sleep apnea in males with an AUC of 0.64 ± 0.03, while MFCC and demographic features led to AUCs of 0.56 ± 0.02 and 0.57 ± 0.02, respectively. Additionally, our results reveal gender-specific patterns in model effectiveness across different medical domains. For males, speaker identification and diarization models consistently outperformed speech foundation models for respiratory conditions (e.g., asthma: 0.61 ± 0.03 vs. 0.56 ± 0.02) and sleep-related conditions (insomnia: 0.65 ± 0.04 vs 0.59 ± 0.05). For females, speaker diarization models performed best for smoking status (0.61 ± 0.02 vs. 0.55 ± 0.02), while Hebrew-specific models performed best (0.59 ± 0.02 vs. 0.58 ± 0.02) in classifying anxiety compared to speech foundation models. Our findings provide evidence that a simple counting task can support large-scale, multi-phenotypic voice screening and highlight which embedding families generalize best to specific conditions, insights that can guide future vocal biomarker research and clinical deployment.
\end{abstract}

\section{Introduction} 
Human speech is a richly layered signal. Beyond lexical content, it carries prosody, articulation, timing, respiration, and other paralinguistic cues that mirror a speaker’s physiology and neurological state\cite{fant1971acoustic,  harma2024survey}. Recent deep learning work has capitalized on these cues for emotion recognition \cite{latif2021survey}, speaker verification \cite{mittal2022automatic}, and detection of neurodegenerative \cite{tao2025early} or pulmonary disorders \cite{sharma2020coswara}.  However, most clinical studies remain narrow: they target a single disease (e.g., Parkinson's disease \cite{moro2021advances} and Alzheimer's disease \cite{luz_detecting_2021}), use small cohorts, or focus on non-speech events such as coughs and breaths \cite{baur2024hear}. 

In this study, we leverage the extensive and uniquely comprehensive data from the Human Phenotype Project (HPP) \cite{shilo_10_2021} to investigate associations between voice characteristics and health conditions. Furthermore, we systematically compare multiple voice feature extraction techniques, ranging from classical spectral features to advanced embeddings\footnote{The terms \textit{embeddings} and \textit{representations} are used interchangeably throughout this paper.} derived from state-of-the-art self-supervised foundation models for speech.  

\textbf{Why fluent counting?} A brief “count-to-30” prompt is easy to administer remotely, highly reproducible, and intuitively understood by speakers of any language \cite{wardle2011quantifying, schobi2022evaluation}. It also elicits sustained phonation and prosodic variability, which are not captured in cough-centric corpora such as Coswara \cite{sharma2020coswara}.

\textbf{The HPP-Voice corpus.}  We introduce \textit{HPP-Voice}, comprising \textbf{7,188 recordings} from 6,760 adults (3,211 males / 3,549 females, age $52\pm10$ years). Each utterance is paired with 15 health conditions spanning across 6 body systems: respiratory, metabolic, neurological, mental health, immune, and sleep \cite{shilo_10_2021}. 

\textbf{Our contributions are:}
\begin{enumerate}
    \item \textbf{Large, clinically diverse corpus.}  We release a novel dataset of fluent speech paired with verified diagnoses across six body systems.
    \item \textbf{Comprehensive benchmark.}  Fifteen state-of-the-art encoders (e.g., MFCC \cite{davis1980comparison}, x-vector \cite{snyder_x-vectors_2018}, wav2vec 2.0 \cite{baevski2020wav2vec}, WavLM \cite{chen2022wavlm}) are evaluated under identical splits.
    \item \textbf{Phenotype-specific insights.} We demonstrate which speech representations outperform for specific clinical domains, providing a comprehensive evaluation of their relative efficacy across different healthcare applications.
\end{enumerate}

\section{Related Work}

Early work on voice-based clinical biomarkers focused primarily on single conditions. These studies relied on handcrafted features related to prosody and phonation to monitor diseases such as Parkinson’s \cite{moro2019study, moro2021advances}, amyotrophic lateral sclerosis (ALS) \cite{suhas2019comparison}, Alzheimer’s disease \cite{luz_detecting_2021}, and various laryngeal disorders \cite{kim2024classification, putzer1997german}. However, such efforts were often constrained by small sample sizes (typically dozens to a few hundred participants), limiting statistical power and the ability to derive insights across conditions or populations.

Recent efforts to scale clinical voice datasets have largely focused on non-speech vocalizations. The COVID-19 pandemic accelerated the creation of cough- and breath-centric corpora, such as Coswara \cite{sharma2020coswara}, MIT-Cough \cite{9208795}, and the HeAR dataset \cite{baur2024hear}, which aggregates recordings from Coswara, CoughVID \cite{orlandic2021coughvid}, and CIDRZ. Although these datasets include thousands of audio samples, they primarily capture explosive airflow events rather than sustained phonation or prosodic variation, thus limiting their utility for analyzing speech patterns.

In parallel, advances in self-supervised learning (SSL) have led to powerful speech representations trained without explicit labels. Models such as Wav2Vec 2.0 \cite{baevski2020wav2vec} and WavLM \cite{chen2022wavlm} now dominate performance benchmarks across a range of tasks \cite{yang_large-scale_2024}, including emotion recognition \cite{wani2021comprehensive, latif2021survey} and speaker identification \cite{jahangir2021speaker}. SSL embeddings consistently outperform traditional baselines such as MFCCs and x-vectors, but prior work has not yet compared such a diverse set of models on a single, clinically annotated speech corpus \cite{baur2024hear}. Efforts to develop voice-based screening for multiple conditions remain rare, with notable exceptions including COPD-Gene Voice \cite{idrisoglu2024copdvd} and HeAR \cite{baur2024hear}, which focus primarily on respiratory traits.

We fill this gap by introducing HPP-Voice and conducting the first head-to-head comparison of fifteen modern encoders on a large, clinically annotated speech corpus for multi-morbid screening.

\section{Dataset: HPP-Voice}
\label{sec:Dataset}
HPP-Voice is the voice arm of the \emph{Human Phenotype Project} (HPP), a longitudinal, multi-omics registry designed to unravel chronic-disease mechanisms at the population scale.

\paragraph{Cohort description}
The HPP cohort currently comprises 11,460 adults. Between December 2019 and December 2024, 6,760 unique speakers were invited for voice sampling and completed the protocol, yielding \textbf{7,188} recordings. Among the 6,760 speakers, 3,549 were women (52\%) aged $52.4\pm10.7$ years and 3,211 were men (48\%) aged $52.6\pm9.9$ years.

\paragraph{Inclusion criteria}
Participants were aged 18–80 years. Individuals with severe or major chronic disorders, such as cancer or Parkinson’s disease, were excluded. All participants signed an informed consent form upon arrival to the research site. All identifying details of the participants were removed prior to the computational analysis. The HPP cohort study is conducted according to the principles of the Declaration of Helsinki and was approved by the Institutional Review Board (IRB) of the Weizmann Institute of Science. Detailed characteristics of the original cohort, including inclusion and exclusion criteria, have been described previously \cite{shilo_10_2021}.

\paragraph{Recording protocol}
Speech was captured in a controlled laboratory environment using 32-bit 384 kHz sampling, then down-sampled to 16 kHz mono. Each participant performed a single 30-second counting task at a comfortable pace. No additional voice tasks were recorded.

\paragraph{Medical labels}
Medical and phenotype labels were derived from the HPP questionnaire and medical records. Sleep apnea was defined as present if the averaged Apnea-Hypopnea Index (AHI) measured across three nights of monitoring exceeded 15 events per hour \cite{kohn_phenome-wide_2025}. Cases were also included if the subject self-reported the condition. All other health conditions were based on self-report. For analysis, we grouped the conditions into six body systems: respiratory (asthma, current smoker, past smoker, chronic sinusitis), metabolic (anemia, hyperthyroidism, hypothyroidism), neurological (migraine, headache), mental health (depression, anxiety), immune (COVID-19, allergy), and sleep-related (sleep apnea, insomnia). Table~\ref{tab:prevalence} presents gender-specific prevalence for each condition.

\paragraph{Data access}
The dataset is available to qualified researchers affiliated with universities or other recognized research institutions upon request at \url{https://humanphenotypeproject.org/home}. Access instructions can be obtained by contacting the project administrators.\footnote{For access inquiries, please contact \texttt{info@pheno.ai}.}

\begin{table}[ht]
\centering
\caption{Label prevalence in HPP-Voice by gender and condition group.}
\label{tab:prevalence}
\begin{minipage}[t]{0.45\textwidth}
\centering
\vspace{0pt}
\begin{tabular}{lcc}
\toprule
\textbf{Condition} & \textbf{Female (\%)} & \textbf{Male (\%)}  \\
\midrule
\multicolumn{3}{l}{\textbf{Respiratory}} \\
Asthma & 4.65 & 5.85 \\
Current smoker & 5.81 & 5.07 \\
Past smoker & 14.09 & 18.60 \\
Chronic sinusitis & 4.17 & 2.30 \\
\midrule
\multicolumn{3}{l}{\textbf{Metabolic}} \\
Hyperthyroidism & 2.14 & 0.72 \\
Hypothyroidism & 9.55 & 2.34 \\
Anemia & 7.41 & 1.90 \\
\midrule
\multicolumn{3}{l}{\textbf{Neurological}} \\
Migraine & 8.26 & 2.90 \\
Headache & 3.10 & 1.31 \\
\bottomrule
\end{tabular}
\end{minipage}%
\hfill
\begin{minipage}[t]{0.45\textwidth}
\centering
\vspace{0pt}
\begin{tabular}{lcc}
\toprule
\textbf{Condition} & \textbf{Female (\%)} & \textbf{Male (\%)}  \\
\midrule
\multicolumn{3}{l}{\textbf{Mental Health}} \\
Depression & 4.20 & 2.55 \\
Anxiety & 3.94 & 2.65 \\
& & \\ 
& & \\ 
\midrule
\multicolumn{3}{l}{\textbf{Immune \& Inflammatory}} \\
COVID-19 & 52.78 & 55.19 \\
Allergy & 20.09 & 18.25 \\
& & \\ 
\midrule
\multicolumn{3}{l}{\textbf{Sleep}} \\
Sleep Apnea & 4.65 & 14.89 \\
Insomnia & 1.94 & 1.18 \\
\bottomrule
\end{tabular}
\end{minipage}
\end{table}

\section{Method}
Our pipeline converts every HPP-Voice recording into a single fixed-length vector, feeds these vectors into downstream classifiers, and compares performance across a diverse set of embedding models.

\subsection{Preprocessing and quality control}

Recordings were first peak-normalized using Librosa~\cite{mcfee_librosalibrosa_2025}, scaling each waveform so its maximum RMS equaled 1. Then, leading and trailing silence was trimmed. 

To identify technical artifacts, a subset of 488 recordings was manually labeled for audio faults. A Random Forest classifier was then trained on this labeled subset and evaluated using 5-fold cross-validation, achieving a mean AUC of $0.95 \pm 0.04$. The classifier was subsequently applied to the full dataset, and recordings with a predicted probability greater than 50\% of being problematic were excluded.

\subsection{Representation families}\label{sec:embeddings}
To compare how different types of acoustic representations capture health-relevant voice information, we systematically evaluate 14 different speech embeddings across five categories, each capturing unique aspects of the speech signal (see Table \ref{tab:embedding-models}). In addition, we compute and evaluate the performance of Mel-Frequency Cepstral Coefficients (MFCCs) as a classical, non-deep learning baseline. Below, we briefly describe and motivate each model:

\paragraph{Speech foundation models}

These self-supervised models learn rich representations from unlabeled speech data through pretext tasks, capturing diverse acoustic and linguistic characteristics without task-specific fine-tuning:

\begin{itemize}
\item \textbf{wav2vec2-Base} \& \textbf{wav2vec2-Large}: These models learn contextual speech representations through contrastive predictive coding. As shown by Baevski et al.~\cite{baevski2020wav2vec}, wav2vec 2.0 employs a quantization module to discretize the latent representations, enabling the model to learn from a contrastive task over masked regions. Yang et al.~\cite{yang_large-scale_2024} demonstrated that the Large variant (317 million parameters) significantly outperforms the base model (95 million parameters) across multiple speech tasks, particularly for content-related tasks like phoneme recognition and ASR.

\item \textbf{WavLM-Base} \& \textbf{WavLM-Large}: WavLM extends wav2vec 2.0 by incorporating a masked speech prediction and denoising framework. According to Chen et al.~\cite{chen2022wavlm}, WavLM uniquely combines masked speech prediction with denoising objectives, allowing it to model both linguistic content and non-ASR information like speaker identity and emotion. Yang et al.~\cite{yang_large-scale_2024} showed that WavLM Large outperforms other models across 14 speech tasks on the SUPERB benchmark~\cite{yang2021superb}, achieving a 2.4-point absolute improvement over previous state-of-the-art, with particular strengths in speaker-related tasks.

\item \textbf{XLSR-53}: As described by Conneau et al.~\cite{conneau2021unsupervised}, XLSR-53 is a cross-lingual extension of wav2vec 2.0 that learns speech representations from raw waveform in multiple languages. It employs a single quantization module shared across languages, which enables effective cross-lingual transfer. Yang et al.~\cite{yang_large-scale_2024} confirmed that this model captures more language-agnostic speech characteristics, showing strong performance even on languages not seen during pre-training.
\end{itemize}

\paragraph{Hebrew-specific models}

Given our Hebrew-speaking cohort, we evaluate language-specific models that leverage ivrit.ai~\cite{marmor_ivritai_2023}, the largest open corpus of Hebrew voice recordings and transcripts dedicated to training Hebrew-specialized speech models.

\begin{itemize}
\item \textbf{XLSR Hebrew-PT}: For this model, we continued the pretraining of XLSR-53 on a subset of recordings from the ivrit.ai corpus, allowing the model to develop robust representations specifically for Hebrew acoustic and phonetic patterns.

\item \textbf{XLSR Hebrew-FT}: We took the XLSR-53 model and fine-tuned it on a Hebrew automatic speech recognition (ASR) task using transcribed data from the Hebrew subset of the Common Voice corpus~\cite{ardila_common_2020}. This approach combines the cross-lingual robustness of XLSR with Hebrew-specific adaptations, which Conneau et al. ~\cite{conneau2021unsupervised} demonstrated is particularly effective for languages not well-represented in the original pretraining data.
\end{itemize}

Both approaches potentially enhance the detection of subtle deviations in Hebrew speech patterns that may correlate with various medical conditions, offering potentially greater sensitivity compared to general multilingual models without Hebrew-specific optimization.

\paragraph{Speaker diarization (SD) models}

SD addresses the challenge of determining "who spoke when" in audio recordings with multiple speakers. These models excel at distinguishing different speakers across diverse acoustic environments:

\begin{itemize}
\item \textbf{pyannote}: Derived from models \cite{bredin_pyannoteaudio_2019} trained on the extensive VoxCeleb corpus~\cite{nagrani2020voxceleb}, this embedding model captures speaker-distinctive characteristics across varying acoustic conditions. It builds upon the x-vector architecture\cite{snyder_x-vectors_2018}, which has shown strong performance in speaker diarization challenges \cite{ryant_third_2021}.

\item \textbf{WavLM-SD}: This embedding leverages the WavLM architecture specifically optimized for SD. Chen et al.~\cite{chen2022wavlm} demonstrated that WavLM achieves state-of-the-art performance on the SUPERB benchmark for speaker diarization tasks, with a relative diarization error rate (DER) reduction of 22.6\% compared to HuBERT Base~\cite{hsu2021hubert}. This superior performance stems from WavLM's masked speech denoising and prediction framework that effectively handles multi-speaker signals during pre-training.
\end{itemize}

\paragraph{Speaker identification (SI) models}

SI models focus on recognizing unique voice signatures that allow for identifying the speaker. These signatures might correlate with one's physiological state:

\begin{itemize}
\item \textbf{x-Vector}: Based on time-delay neural networks (TDNNs) with statistical pooling~\cite{ravanelli_speechbrain_2021}, this embedding has demonstrated excellent performance in extracting speaker-specific characteristics. The x-vector architecture has been at the forefront of speaker recognition research \cite{snyder_x-vectors_2018}.

\item \textbf{EfficientNet}: We trained this model from scratch specifically on our corpus (HPP-Voice) using a contrastive learning approach, where the model learns to align segments from the same recording while distinguishing them from segments of other recordings. For this purpose, we employed the computationally efficient EfficientNet architecture \cite{tan2019efficientnet} as the encoder for voice segments. This self-supervised training strategy allows the model to learn speaker-specific characteristics (i.e., model embeddings) without relying on explicit speaker labels, while retaining the parameter efficiency inherent to the EfficientNet design.

\item \textbf{pyannote-FT}: For this model, we used the \textit{pyannote} model (pretrained on the VoxCeleb corpus) for segment encoding and fine-tuned it specifically for speaker identification in our corpus (HPP-Voice), following the same contrastive framework described for \textit{EfficientNet} above. This approach was aimed at optimizing the model's ability to capture distinctive voice characteristics within our study population.

\end{itemize}

\paragraph{Emotion-specific models}

Emotion models detect affective states through vocal cues, which may be relevant for both psychological and physiological condition detection:

\begin{itemize}

\item \textbf{wav2vec2-SER}: Based on the wav2vec2 architecture but optimized for Speech Emotion Recognition (SER) through supervised fine-tuning on emotional speech corpora. As detailed by Ravanelli et al.~\cite{ravanelli_speechbrain_2021}, this model targets the paralinguistic features that may correlate with both affective disorders and physical conditions that influence voice modulation.

\item \textbf{WavLM-SED}: This model extends WavLM with emotion diarization capabilities. Wang et al.~\cite{wang_speech_2023} introduced the concept of Speech Emotion Diarization (SED), where the model identifies not just what emotions are present but precisely when they occur in an utterance.
\end{itemize}

\paragraph{Mel‑Frequency Cepstral Coefficients (MFCC)}
A standard front‑end in speech analysis, MFCCs are usually truncated to the first $13$ coefficients, which capture the bulk of the vocal‑tract spectral envelope while keeping dimensionality low~\cite{hibare_feature_2014}.
The resulting compact, noise‑robust vector is widely adopted for detecting physiologically driven voice changes across clinical and paralinguistic tasks~\cite{tracey_towards_2023}.

\begin{table}[ht]
\caption{Speech-embedding models evaluated in this study. Dataset acronyms: LS (LibriSpeech)~\cite{panayotov2015librispeech}, LL (LibriLight)~\cite{kahn2020libri}, LVG (94k-hour dataset of LibriLight, VoxPopuli~\cite{wang2021voxpopuli}, and Gigaspeech~\cite{chen2021gigaspeech}). Process acronyms: PT (Pre-training), FT (Fine-tuning). Task acronyms: SD (Speaker Diarization), SI (Speaker Identification), SER (Speech Emotion Recognition), SED (Speech Emotion Diarization), ASR (Automatic Speech Recognition).}
  \label{tab:embedding-models}
  \centering
  \begin{tabular}{p{2.5cm}p{3.1cm}p{1.5cm}p{5.2cm}}
    \toprule
    \textbf{Family} & \textbf{Model} & \textbf{\# Params} & \textbf{Training Data} \\

    \midrule
    \multirow{5}{*}{Speech foundation} & wav2vec2-Base~\cite{baevski2020wav2vec}          & 95 M & PT: LS 960 hr \\
    & wav2vec2-Large~\cite{baevski2020wav2vec}         & 317 M & PT: LL 60k hr \\
    & WavLM-Base~\cite{chen2022wavlm}             & 95 M & PT: LS 960 hr \\
    & WavLM-Large~\cite{chen2022wavlm}                & 317 M & PT: LVG 94k hr \\
    & XLSR-53~\cite{conneau2021unsupervised}               & 300 M & PT: 53-lang mixed corpus 56k hr \\
    \midrule
    \multirow{2}{*}{Hebrew} & XLSR Hebrew-PT      & 300 M & PT: XLSR-53 corpus + ivrit.ai  \\
    & XLSR Hebrew-FT      & 300 M & PT: same as above; FT: Hebrew ASR on Common Voice/he 8 hr subset \cite{ardila_common_2020} \\
    \midrule
    \multirow{2}{*}{SD} & WavLM-SD~\cite{chen2022wavlm}     & 317 M & PT: LVG 94k hr; FT: SD on LibriMix \\
    & pyannote~\cite{bredin_pyannoteaudio_2019} & 4.3 M & PT: VoxCeleb~\cite{nagrani2020voxceleb} \\
    \midrule
    \multirow{3}{*}{SI} & pyannote-FT            & 4.3 M & PT: VoxCeleb; FT: SI on HPP-Voice \\
    & x-vector~\cite{snyder_x-vectors_2018}        & 4.2 M & PT: VoxCeleb \\
    & EffNet~\cite{tan_efficientnet_2020}      & 6 M & PT: HPP-Voice \\
    \midrule
    \multirow{2}{*}{Emotion} & wav2vec2-SER~\cite{ravanelli_speechbrain_2021}            & 95 M & PT: LS 960 h; FT: Emotion recognition on IEMOCAP~\cite{busso2008iemocap}  \\
    & WavLM-SED~\cite{wang_speech_2023}     & 317 M & PT: LVG 94k hr; FT: subsets from English emotion datasets, e.g., IEMOCAP, RAVDESS~\cite{livingstone2018ryerson} \\
    \bottomrule
  \end{tabular}
\end{table}

\subsection{Experimental setup}

\paragraph{Embedding extraction}
Each model outputs either (i) one embedding per recording or (ii) frame-level $d$-dimensional vectors. For frame-level outputs, we apply simple mean pooling over time to obtain a single $1 \times d$ representation, ensuring that all recordings share an identical feature dimensionality regardless of model family. To standardize input lengths and reduce variability, recordings were segmented into 5-second-long segments, and each segment was embedded separately.

\paragraph{Gender-specific stratification based on confounding analysis}

Preliminary testing for potential confounding variables revealed that all evaluated embedding models were able to predict participant gender with very high performance, achieving AUCs ranging from 0.92 (MFCC) to 0.98 (WavLM). These results are consistent with prior work demonstrating that gender can be reliably inferred from voice characteristics~\cite{kwasny_gender_2021}. To better assess the suitability of each embedding model for detecting specific health phenotypes, we stratified the dataset into male and female subsets prior to model training and evaluation. After applying quality control steps, removing repeated visits, we retained 2,150 recordings from unique female participants and 1,993 recordings from unique male participants.

\paragraph{Classifier training and evaluation}

We employed a LightGBM classifier~\cite{ke_lightgbm_2017} trained using 4-fold cross-validation, with hyperparameter optimization conducted via Optuna~\cite{akiba_optuna_2019} using 20 trials per fold. For each combination of embedding model, target phenotype, and gender group, a separate classifier was trained. To ensure robustness and reduce variance due to stochastic effects, the entire process was repeated across 20 random seeds. For classifier training and evaluation, only the first embedded segment, corresponding to the first 5 seconds after trimming leading silence, was used.

The baseline model was trained using the demographic confounder age as the sole input feature. For all other embedding models, age was included as an additional input feature to adjust for its potential confounding effect during training.

To evaluate whether embedding-based models achieved statistically significant improvements over the age-only baseline, we performed pairwise comparisons across seeds using the Wilcoxon signed-rank test. Multiple hypothesis testing was corrected using the Benjamini–Hochberg procedure, with a significance threshold of \( q < 0.05 \).

\paragraph{Computational infrastructure}
The XLSR Hebrew pretraining was conducted on an NVIDIA Quadro RTX-8000 GPU for 12 hours. Embedding extraction for all models was performed on the same GPU configuration, requiring less than 10 hours of total computation time. LightGBM classifier training and evaluation were executed on AMD EPYC™ 7002 24-core and Intel Xeon® Gold 6140 18-core @ 2.30GHz processors. Each model utilized 10 CPU cores during training, with hyperparameter optimization and evaluation requiring approximately 6 hours per random seed across all 20 repetitions.

\section{Results}
\label{sec:Results}

\begin{figure}[htbp]
\centering
\includegraphics[width=0.99\textwidth]{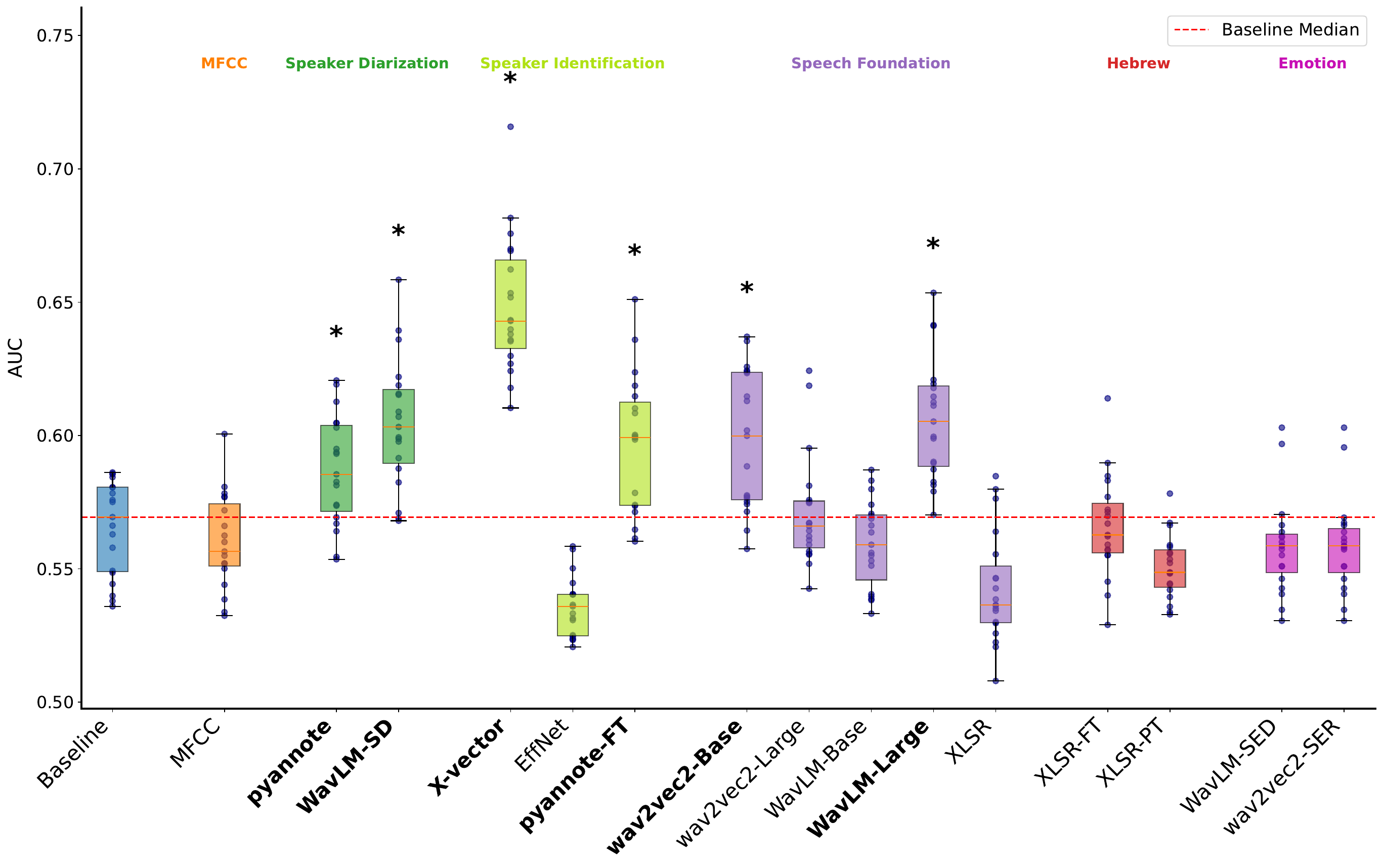}
\caption{\textbf{Sleep apnea detection performance in males.} Performance comparison of speech representation models for sleep apnea detection in males. Models are grouped by speech processing domain (color-coded at top). Boxplots show AUC distribution across 20 random seeds. Asterisks (\textbf{*}) indicate statistically significant improvement over baseline (Wilcoxon signed-rank test; FDR-BH corrected).}
\label{fig:SA_comparison}
\end{figure}

We begin our analysis with sleep apnea, as this condition, unlike the others in our study that rely on self-reported questionnaires, was curated using objective physiological recordings. Specifically, condition labels were derived from multi-night measurements using Itamar Medical devices that monitor chest movements, pulse rate, peripheral blood, oxygen saturation, sleep stages, body positions, and snoring patterns, assessing clinical-grade sleep apnea diagnosis.

Our analysis focused on sleep apnea prediction in males, where this disorder is more prevalent (14.89\% in our cohort). As shown in Figure~\ref{fig:SA_comparison}, speaker identification models demonstrated superior performance in detecting moderate to severe sleep apnea. The x-vector model achieved the highest performance with an AUC of $0.64 \pm 0.03$, significantly outperforming both MFCC features (AUC = $0.56 \pm 0.02$) and baseline demographic features (AUC = $0.57 \pm 0.02$). The pyannote-FT model also showed significant improvement over the baseline. Among speech foundation models, WavLM-Large exhibited strong performance with significant gains over baseline, while models trained for emotion recognition (WavLM-SED, wav2vec2-SER) and language-specific models (XLSR Hebrew-FT, XLSR Hebrew-PT) did not significantly improve over baseline.

\begin{figure}[htbp]
\centering
\includegraphics[width=\textwidth]{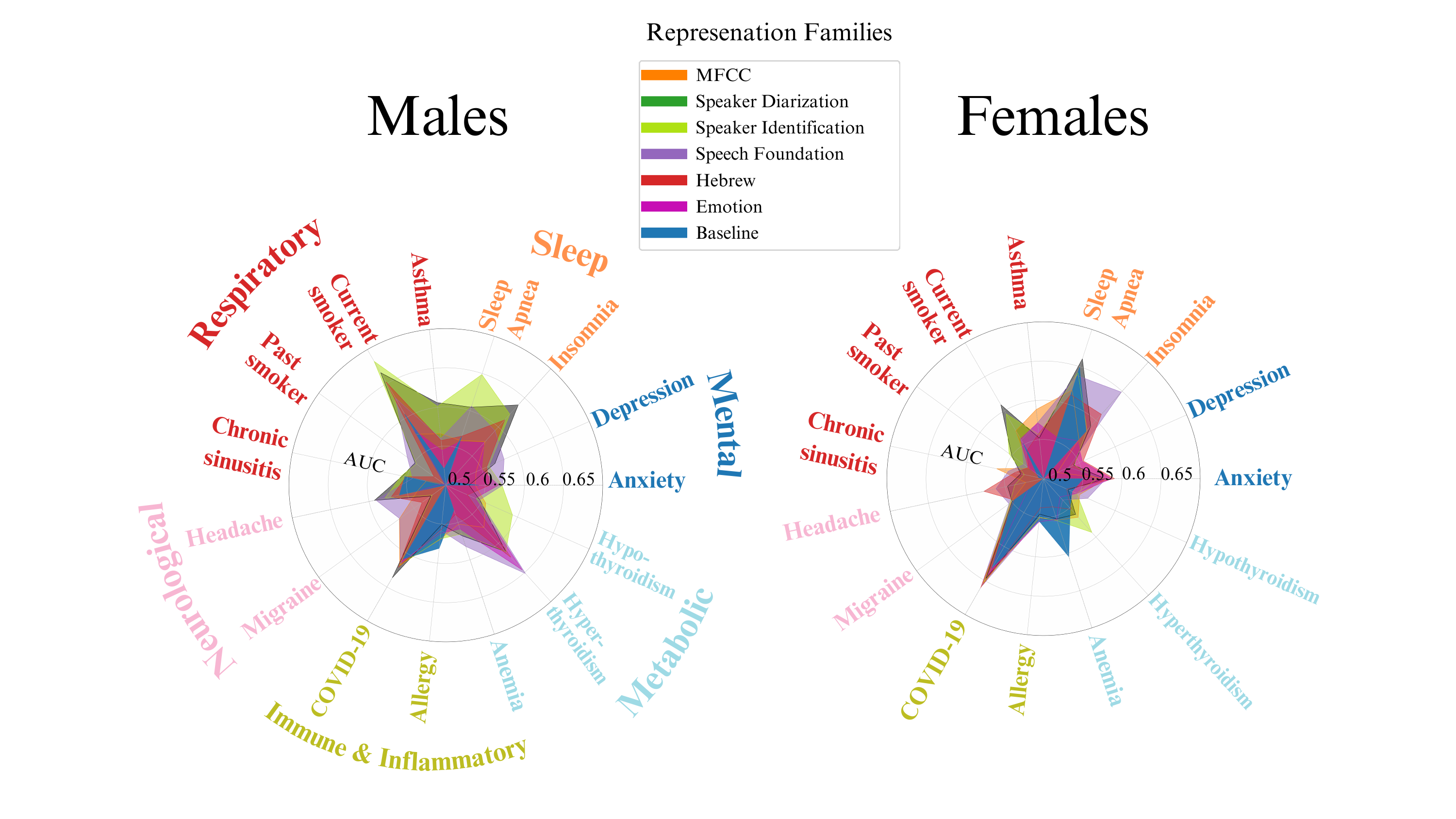}
\caption{\textbf{Gender-specific model performance across medical domains.} Radar plots showing average AUC scores of the best-performing model from each speech representation family across medical conditions grouped by domain. The left panel shows male subjects; the right panel shows female subjects. The plots demonstrate distinct gender-dependent patterns in model performance across various medical conditions.}
\label{fig:radar}
\end{figure}

Beyond sleep apnea, we extended our analysis to multiple self-reported medical conditions, as shown in Figure~\ref{fig:radar}. The radar plots display the performance of the best model from each speech representation family across diverse clinical domains. The analysis reveals pronounced gender-specific patterns in model effectiveness. For males, SI and SD models consistently outperformed speech foundation models for respiratory conditions (e.g., asthma: $0.61 \pm 0.03$ vs. $0.56 \pm 0.02$) and sleep-related conditions (insomnia: $0.65 \pm 0.04$ vs. $0.59 \pm 0.05$). In contrast, among females, SD models performed best for smoking status ($0.61 \pm 0.02$ vs. $0.55 \pm 0.02$), while Hebrew-specific models achieved superior performance ($0.59 \pm 0.02$ vs. $0.58 \pm 0.02$) in classifying anxiety compared to speech foundation models.
A comprehensive evaluation of all tested medical conditions is provided in the Supplementary Material.

These results demonstrate that different speech representation families exhibit varying levels of effectiveness depending on both the specific medical condition and the gender of the speaker, suggesting that no single model architecture is optimal across all contexts.
\section*{Discussion and limitations}

This study introduces HPP-Voice, a large-scale corpus of Hebrew speech paired with multi-system health phenotypes, and systematically benchmarks 14 speech embedding models for the task of multi-condition classification. Using only a 30-second counting task per subject, we demonstrate that modern speech representations, particularly those trained for speaker identification and diarization, can detect clinically relevant signals for a variety of conditions, including sleep apnea, asthma, smoking status, and anxiety, accounting for age and gender as potential confounders. The optimal embedding family varied by both medical domain and gender, indicating the presence of condition-specific and population-specific acoustic markers.

Several hypotheses may explain why counting, despite its constrained linguistic content, encodes such health-relevant information. First, the dynamics of breathing pauses during fluent counting may reflect respiratory health and sleep-disordered breathing patterns, as suggested by work using breath and cough sounds for diagnosis \cite{sharma2020coswara}. Second, prosodic features such as pitch variability, rhythm, and speech rate are known to reflect psychological and neurological states \cite{luz_detecting_2021, latif2021survey}. Third, articulatory stability and vocal tract control, which are often captured by MFCCs and speaker models, may be modulated by metabolic or cognitive impairments \cite{tracey_towards_2023}. These paralinguistic cues are naturally captured by embeddings optimized for speaker identity or diarization tasks, which may explain their strong performance for physical health conditions.

Our analysis also reveals that gender-specific performance patterns emerge across medical domains. For example, speaker diarization and identification models were more effective in detecting respiratory and sleep disorders in males, whereas Hebrew-specific models improved the classification of anxiety in females. This supports the hypothesis that vocal manifestations of health conditions differ across demographic groups and that task-specific or population-specific model tuning may be essential for clinical translation.

This work has several limitations. All recordings were made using studio-grade microphones in a controlled environment, which may not reflect the variability present in mobile or telehealth deployments \cite{baur2024hear}. Each participant contributed only a single utterance, precluding analysis of temporal dynamics or disease progression. The dataset is restricted to Hebrew-speaking adults, limiting its generalizability to other linguistic and cultural populations \cite{conneau2021unsupervised, marmor_ivritai_2023}. While sleep apnea labels were derived from objective multi-night physiological monitoring, most other health conditions were based on self-report \cite{shilo_10_2021}, which introduces potential label noise. Despite these constraints, the ability of several models to outperform demographic and MFCC baselines suggests that vocal biomarkers can be robust to such sources of variation.

Future research should aim to address these limitations through longitudinal tracking of vocal changes, in-the-wild mobile data collection, and multilingual corpus expansion. These steps will be crucial to assessing real-world utility and fairness across diverse populations. Further work is also needed to dissect which acoustic dimensions are most informative for different conditions and to develop interpretable, condition-specific models.

Nonetheless, our findings underscore the clinical potential of voice as a scalable, low-effort health monitoring modality. A short counting task, easily administered in remote settings, may support non-invasive pre-screening for multiple health conditions, particularly when integrated into telemedicine platforms or deployed in underserved populations. By identifying which embedding families generalize best to specific conditions, this work offers practical guidance for future development of voice-based diagnostic systems.

\section{Conclusion}

This work demonstrates that a simple counting task can serve as a powerful source of paralinguistic signals for multi-condition health screening. Using the HPP-Voice corpus, we benchmarked 14 diverse speech embedding models and identified strong condition- and gender-specific patterns in predictive performance. Speaker identification and diarization embeddings proved most effective for physiological conditions, while language-specific models improved detection of psychological traits. These findings underscore the potential of voice as a scalable, non-invasive modality for remote health assessment and offer practical guidance for model selection in future clinical applications.

\bibliographystyle{plainnat}
\bibliography{references, references-2}

\end{document}